\documentclass[conference]{IEEEtran}

\usepackage[english]{babel}
\usepackage{multirow}
\usepackage{booktabs} 
\usepackage{subcaption}
\usepackage{microtype}
\usepackage{pgfplotstable} 
\usepackage{hyphenat}
\usepackage[ruled,vlined]{algorithm2e}
\usepackage{setspace} %
\pgfplotsset{compat=newest}
\pgfplotstableset{fixed zerofill, sci zerofill, precision=2}
\pgfkeys{/pgf/number format/.cd,std=-5:4,precision=2}
\pgfkeys{/pgf/number format/sci e}
\pgfkeys{/pgf/number format/sci generic =  {mantissa sep=e{#1}}}
\pgfkeys{/pgf/number format/sci precision=2}

\SetKwProg{Fn}{Function}{}{end}
\SetKwProg{Str}{Structure}{}{end}

\newcommand{\lfid}{LFID\xspace} %
\newcommand{\figspace}{-0.46em}

\begin{document}
\title{
Hop-by-Hop Multipath Routing:\\ Choosing the Right Nexthop Set
}
\author{\IEEEauthorblockN{Klaus Schneider}
\IEEEauthorblockA{University of Arizona\\
Email: klaus@cs.arizona.edu}
\and
\IEEEauthorblockN{Beichuan Zhang}
\IEEEauthorblockA{University of Arizona\\
Email: bzhang@cs.arizona.edu}
\and
\IEEEauthorblockN{Lotfi Benmohamed}
\IEEEauthorblockA{NIST\\
Email: lotfi.benmohamed@nist.gov}}
\maketitle

\begin{abstract}
The Internet can be made more efficient and robust with hop-by-hop multipath routing:
Each router on the path can split packets between multiple nexthops in order to 1) avoid failed links and 2) reduce traffic on congested links.
Before deciding \emph{how} to split traffic, one first needs to decide which \emph{nexthops} to allow at each step.
In this paper, we investigate the requirements and trade-offs for making this choice.

Most related work chooses the viable nexthops by applying the ``Downward Criterion'', i.e., only adding nexthops that lead closer to the destination; 
or more generally by creating a Directed Acyclic Graph (DAG) for each destination.
We show that a DAG's nexthop options are necessarily limited, and that, by \emph{using certain links in both directions} (per destination), we can add further nexthops while still avoiding loops.
Our solution \lfid (Loop-Free Inport-Dependent) routing,
though having a slightly higher time complexity, leads to both a \emph{higher number} of and \emph{shorter} potential paths than related work.
\lfid thus protects against a higher percentage of single and multiple failures (or congestions) and comes close to the performance of arbitrary source routing.

\end{abstract}

\section{Introduction}

While traditional routing protocols (like OSPF \cite{ospf_v2_rfc2328} or IS-IS \cite{isis_rfc1142}) send packets on the \emph{shortest path} from source to destination, there are now many recognized benefits of using multiple \emph{non-shortest paths} \cite{rexford2008toward}, most importantly 1) \emph{failure protection:} routing around link/node failures, and 2) \emph{traffic engineering:} distributing the traffic load to avoid congestion. 

A common approach to failure protection is \emph{IP Fast Rerouting} (IPFRR) \cite{ipfrr_rfc}, which provides alternative nexthops for quick local packet detours on the data plane.
Compared to shortest path routing, IPFRR has clear benefits:
it can reroute packets almost instantly instead of waiting for routes to converge (often hundreds of milliseconds), which avoids loops and packet drops during the convergence phase \cite{ipfrr_rfc}.

A common approach to traffic engineering is to use end-to-end (E2E) tunnels where the source (ingress) node determines the entire path towards the destination, as done by MPLS \cite{mpls_rfc3031} or segment routing \cite{segment_routing_rfc8402}. 
Though frequently used in practice, E2E traffic engineering does have certain downsides: 
1) Scalability: endpoints need to select a small number of actual paths from an exponential number of potential ones, trading off path diversity and path length. 
2) Data plane overhead: packets need to carry additional headers to steer them through the network.

\begin{figure}
\footnotesize
\begin{subfigure}{0.4\linewidth}
\begin{tabular}{@{}llcr@{}}
\toprule
\textbf{Dest.} &  \textbf{NH} & \textbf{\hspace*{-.3em}Cost\hspace*{-.4em}} & \textbf{Split} \\ 
\midrule
LA 											& SV	& 2 & 100\%  \\
\midrule
\multirow{2}{0em}{DV}   & DV	& 1 & 100\%  \\
									    	& SV	& 2 &   0\%  \\
\midrule
\multirow{2}{0em}{NYC}  & DV	& 5 &  60\%  \\
									    	& SV	& 6 &  40\%  \\
\midrule
\ldots \\ 
\bottomrule
\end{tabular}
\caption{FIB for \textbf{Seattle}}
\label{fig:fib_example_a}
\end{subfigure}
\begin{subfigure}{.58\linewidth}
\includegraphics[width=\linewidth]{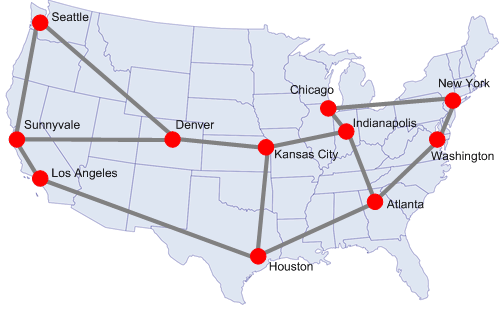}
\caption{Abilene Topology}
\label{fig:fib_example_b}
\end{subfigure}

\caption{Ex. of HBH Multipath Routing. Depending on the destination, Seattle uses either 1 or 2 nexthops. The cost denotes the shortest path (in hops) of using a certain nexthop. The split ratio can be adapted to react to link failures or congestion.
\vspace*{\figspace}}
\label{fig:fib_example}
\end{figure}

In this work, we consider an approach to combine both failure protection and traffic engineering: 
\emph{Hop-by-Hop} (HBH) Multipath Routing,
which shares the benefits of IP Fast Rerouting (instant failure protection without relying on routing convergence)
and also provides traffic engineering without per-packet overhead by distributing the path/nexthop decision throughout the network.
In HBH Multipath Routing, each router's FIB is equipped not with a single nexthop per destination, but with a set of nexthops.
Every router on the path (not just edge routers) can then steer traffic by changing the \emph{split ratio} for each of its \emph{nexthops} (not for the entire path) -- see an example in Figure \ref{fig:fib_example}. 
These nexthops can be used both for failure protection (instantly reset the split ratio to 0\%) and for reducing congestion (gradually change the split ratio).

HBH Multipath Routing comprises two fundamental steps: 
First, one needs to decide \emph{which} nexthops to put in the FIB and what cost these nexthops should be assigned. This step determines the \emph{potential} paths that packets can take, based on the network topology and link cost/propagation delay.
Second, one needs to determine the \emph{split ratio} for each nexthop. %
This step selects the \emph{actual} paths from the vast number of potential ones, considering the current network state, such as link failures and link utilization. 
In this paper, we focus solely on the \emph{first step:} choosing the right nexthop set. 

To choose the right nexthop set, we first establish a list of requirements. 
These requirements overlap with the ones for pure failure protection (see IPFRR in Section \ref{sec:related_work_back}), but with two crucial differences:
First, \emph{path lengths are more important:} 
For failure protection, backup paths are used only temporarily until either the failed link is restored or the routing protocol finds a better path -- maintaining connectivity is more important than optimality. 
But for HBH traffic engineering, multiple paths may be used for a much longer duration, making it crucial to maintain a short length.
Second, \emph{failed links are bidirectional, congestion is unidirectional.} Thus, if router A detects that link (A$\rightarrow$B) is down, it is safe to assume that link (B$\rightarrow$A) is also down, which extends A's rerouting options.
As a corollary of this, one can use the incoming port of packets to infer which other links have failed, and thus to alter forwarding decisions \cite{uturn-rfc-draft,vo2013routing,yang2014keep,quest2016}. 
However, when splitting traffic to avoid congestion, other nodes' forwarding options are naturally less restricted (due to the absence of failures), thus nexthops must be chosen \emph{more} restrictively:

\begin{enumerate}
\item \textbf{Avoid loops for arbitrary NH choice:} Packets must not return to a node they have previously visited, even if any number of routers can choose independently (without communication) and arbitrarily from their nexthop set.
An example violation of this requirement is shown for \emph{Loop-free Alternates} in Section \ref{sec:related_work_back}.
  
\item \textbf{No per-packet state:} No packet-specific state is manipulated in either packet headers or the FIB. 
Thus, forwarding decisions are based only on a packet's destination, incoming port, and currently failed links.

\item \textbf{High Number of NHs:} A high number of nexthops (and thus potential paths) allows routers to circumvent more cases of link/node failure and congestion.

\item \textbf{Short Paths:} The potential paths resulting from the nexthop set should be kept as short as possible.
This helps to use the available link capacity more efficiently, and also to reduce the end-to-end latency for delay-sensitive traffic, like audio/video conferencing.

\item \textbf{Preserve accurate path length in the FIB:} The data plane should be given an accurate estimate of the path length caused by using a certain nexthop, to increase traffic engineering efficiency. %
We show the downsides of violating this requirement when discussing ECMP link-weight tuning in Section \ref{sec:inc_nh_choice}.
\end{enumerate}

Existing HBH Multipath Routing schemes \cite{narvaez1999efficient, ospf-omp-02, xu2007deft, xu2011link, gojmerac2008towards, paganini2009unified, michael2015halo, yi2013case}) have, often implicitly, answered these requirements with a natural solution called the ``Downward Criterion'': Routers only add nexthops that lead closer to the destination. 
More generally, the idea is to turn the network into a \emph{directed acyclic graph} (DAG) per destination, where arcs in the graph represent viable nexthops (see Figure \ref{fig:abilene_routings}).

We find that downward paths and DAGs satisfy requirements 1), 2) and 5), but  often fall short in the number of potential paths and sometimes lead to longer paths than necessary (see Section \ref{sec:inc_nh_choice} and \ref{sec:evaluation}).
In our work, \lfid, we extend the concept of DAGs to use certain links in \emph{both directions} (per destination), but we exhaustively prune nexthops so that the remaining paths are guaranteed to be loop-free (= acyclic) when excluding the incoming port at each step. 

Our contribution, \lfid, can be interpreted in two ways: 
First, it is a \emph{local failure rerouting} scheme that, without using per-packet state, gives close to optimal protection against an arbitrary number of uncorrelated link or node failures. 

Second, it is a \emph{first step towards HBH traffic engineering.} 
Now using a small amount of state in routing tables (the \emph{split ratio}) in order reduce the load on congested links. 
Here \lfid only specifies the first step: Which nexthop set to choose, to create good \emph{potential} paths. 
It leaves the rest to future work, e.g., how to detect congestion, what granularity to split (per-flow/per-prefix), or how exactly to determine the split ratio.
However, irrespective of these specifics, we show that \lfid comes close to the optimal, in number of provided paths and with an average path stretch of only +1\% (Section \ref{sec:ev_mult_failures}).
Compared to DAG-based work, \lfid protects against a
higher percentage of single and multiple failures/congestions.
Moreover, \lfid often provides better protection at the node \emph{adjacent} to the failure than related work can by \emph{backtracking} to earlier nodes or all the way to the source (see Section \ref{sec:ev_single_failure}).
Lastly, \lfid does have a slightly higher time and space complexity than related work (Section \ref{sec:design_complexity}), but we show it to be scalable for networks with at least multiple hundreds of routers (Section \ref{sec:ev_time}).
\lfid can be implemented as an extension of current link-state routing; all required topology information is already signaled in a link-state protocol like OSPF. The only changes made are to the route calculation part.

\section{Increasing the Nexthop Choice}
\label{sec:inc_nh_choice}

We first discuss related work that meets at least the first two requirements: being loop-free when routers arbitrarily choose nexthops, without using per-packet state. 
Later in Section \ref{sec:related_work_back}, we discuss work
in the area of IP Fast Rerouting (IPFRR) which either requires per-packet state, or restricts nexthop choice, for example, to only use backup nexthops once all primary nexthops are down.

The earliest and simplest of the related work is \emph{Equal Cost Multi-Path} (ECMP) routing \cite{ecmp_rfc2992}, in which a router uses all nexthops that share the exact shortest path cost towards the destination\footnote{We use the terms ``path cost'' and ``path length'' interchangeably. Similarly, we use the terms ``link cost'',  ``link weight'', and ``link metric'' interchangeably.}.
ECMP paths are always loop-free and as short as possible.
However, the constraint of equal cost matching creates a dilemma: the more \emph{fine-grained} the link cost metric, the less nexthops will be available.
For example, using the inferred link weights of the Rocketfuel topologies (ranging from 1 to 22.5, in steps of 0.5), only 9.7\% to 16.8\% of nodes can protect against the failure of an adjacent link (see Figure \ref{fig:eval_resilience_single} in Section \ref{sec:evaluation}).
If the link metric is set to the hop count these numbers rise to 29.4\% to 59.4\%, but this ignores the real cost of paths (e.g., determined by physical distance), and still provides less protection than the work discussed below. 
Lastly, some work \cite{fortz2000internet, fortz2002traffic, iannaccone2004feasibility} suggests to further increase the number of equal-cost paths by carefully tuning the link weights to that goal. 
This does increase the nexthop choice, but also violates requirement 5 (preserving the real path cost), with undesirable results: 
Now, the forwarding plane treats some paths of different length as equal, which leads to inefficient use of network resources and higher end-to-end delay. 
\begin{figure}
\footnotesize
\begin{subfigure}{.315\linewidth}
\includegraphics[width=\linewidth]{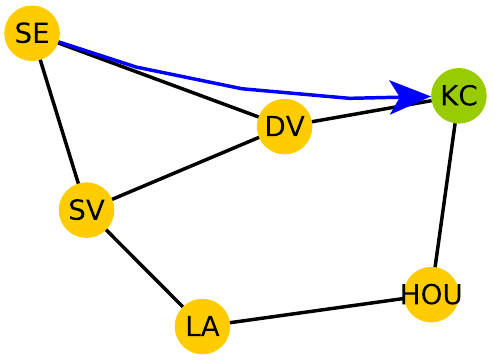}
\caption{}
\label{fig:ecmp_tuning_base}
\end{subfigure}
\hfill
\begin{subfigure}{.315\linewidth}
\includegraphics[width=\linewidth]{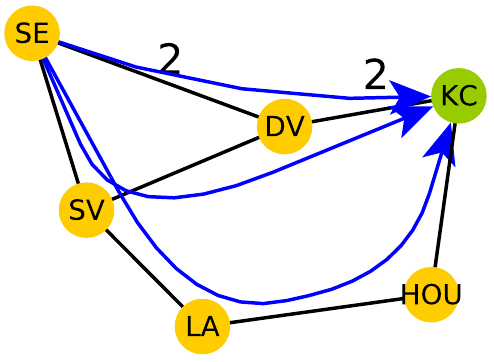}
\caption{}
\label{fig:ecmp_tuning_manip}
\end{subfigure}
\hfill
\begin{subfigure}{.33\linewidth}
\centering
\begin{tabular}{@{}llcr@{}}
\toprule
\textbf{Node\hspace*{-.6em}} & \textbf{NH} & \textbf{\hspace*{-.5em}Cost\hspace*{-.6em}} & \textbf{Split} \\ 
\midrule
\multirow{2}{0em}{SE} 	& DV	& 4 &	33\%  \\
		 										& SV	& 4 &	66\%  \\
\midrule
\multirow{2}{0em}{SV}   & DV	& 3 &	50\%  \\
									    	& LA	& 3 &	50\%  \\
\bottomrule
\end{tabular}
\caption{}
\label{fig:ecmp_tuning_fib}
\end{subfigure}
\caption{Example of ECMP + Link weight tuning. a) Base topology that allows only one equal-cost path. b) Topology with changed link weights to create three equal-cost paths. c) FIB for changed topology with destination KC. 
\vspace*{\figspace}}
\label{fig:ecmp_tuning}
\end{figure}
Consider the example in Figure \ref{fig:ecmp_tuning} for traffic from Seattle to Kansas City. If the link metric is set to the hop count (i.e., every link has a weight of 1), ECMP will only provide one path: SE$\rightarrow$DV$\rightarrow$KC (Figure \ref{fig:ecmp_tuning_base}). 
Now, one can adjust the link weights to add a path SE$\rightarrow$SV$\rightarrow$DV$\rightarrow$KC, or even SE$\rightarrow$SV$\rightarrow$LA$\rightarrow$HOU$\rightarrow$KC, as done in Figure \ref{fig:ecmp_tuning_manip}.
However, the only way ECMP can use these paths is to split traffic on them \emph{equally}, i.e., every path receives about 33\% of traffic (Figure \ref{fig:ecmp_tuning_fib}).
As long as link utilization is below capacity, this is wasteful: some traffic that could have used the shortest path (SE$\rightarrow$DV$\rightarrow$KC) is needlessly sent over a longer path.
A better approach would be to keep traffic on the shortest path until demand exceeds its capacity, and only then switch to the longer path.
However, ECMP + link weight tuning cannot achieve this, as it doesn't preserve the real cost of the path.

A higher path choice and a more accurate cost representation
can be reached by \emph{non-equal cost} multipath algorithms, the most prominent of which is the \emph{Downward Path Criterion} (DW) \cite{rexford2008toward}. 
Downward paths relax the equal cost constraint by including the shortest path nexthop plus any nexthop $n_i$ that is closer to the destination (has a lower cost) than the current node $x$:
$cost(n_i) < cost(x)$. %
Downward nexthops are simple to compute, requiring only one shortest path computation for each neighboring node, and achieve a higher path choice than ECMP. %
Thus, they are used frequently in the literature, known under the names of Loop-Free Invariant (LFI) \cite{vutukury1999simple}, ``viable'' nexthops \cite{narvaez1999efficient}, Rule 1 (One Hop Down) Deflection Set \cite{source_selectable_2006}, and Relaxed Best Path Criterion \cite{ospf-omp-02}. 

One extension of Downward Paths (which we'll call downward+equal -- DWE) is to also consider nexthops with the same cost: $cost(n_i) \leq cost(x)$.
However, to prevent packets from forming loops, one needs to add a \emph{tiebreaker} which assures that traffic only crosses one direction of the equal-cost link. This tiebreaker could be based on the node degree \cite{kvalbein2009multipath} or simply the node id: 
$cost(n_i) < cost(x) \  \vee	 
(cost(n_i) = cost(x) \wedge id(n_i) < id(x))$.
This approach is still guaranteed to be loop-free and, compared to downward paths, provides at least as many nexthops, and often more. 

A further improvement of nexthop choice is achieved by the three algorithms from the work ``Maximum Alternative Routing Algorithm'' (MARA) \cite{mara2009}, which create a ``Maximum Adjacency Ordering'', that is equivalent to turning the network into Directed Acyclic Graphs (DAGs). 
Most relevant for our purposes are the variants MARA-MC, which ``maximizes the minimum node connectivity'', and MARA-SPE which does the same but with the constraint to always include the shortest path tree (see Section \ref{sec:evaluation}).

\begin{figure}
\footnotesize
\begin{subfigure}{0.45\linewidth}
\includegraphics[width=0.4\linewidth]{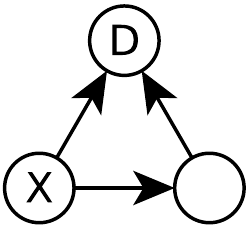}
\includegraphics[width=0.49\linewidth]{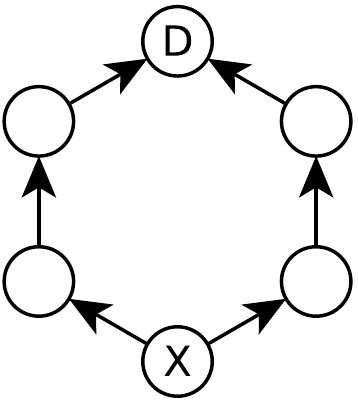}
\caption{DAG}
\label{fig:triangle_ring_dag}
\end{subfigure}
\hfill
\begin{subfigure}{0.45\linewidth}
\includegraphics[width=0.4\linewidth]{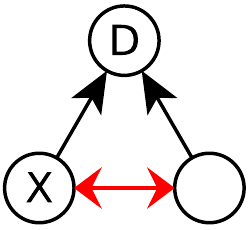}
\includegraphics[width=0.49\linewidth]{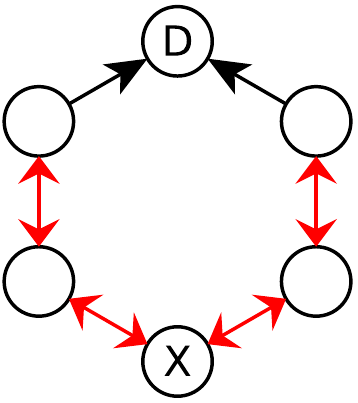}
\caption{\lfid}
\label{fig:triangle_ring_lfip}
\end{subfigure}
\caption{Triangle and Ring topologies
\vspace*{\figspace}}
\label{fig:triangle_ring}
\end{figure}

\begin{figure*}
\begin{subfigure}{0.326\textwidth}
	\includegraphics[width=\linewidth]{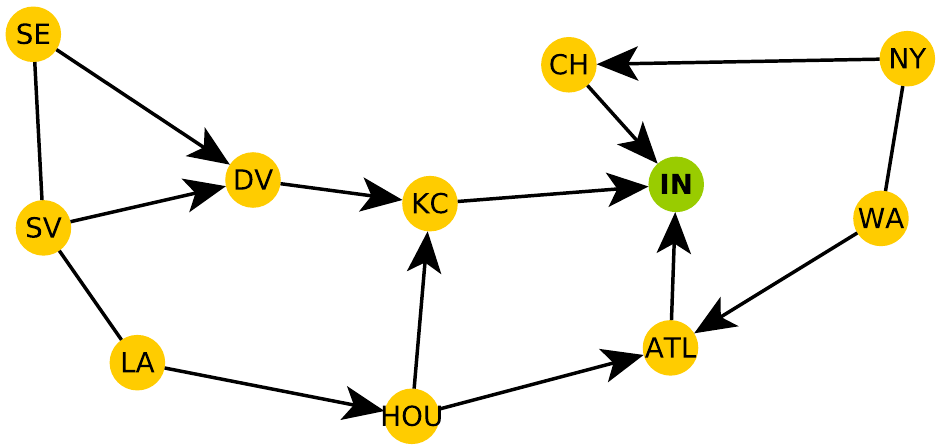}
	\caption{Downward Paths}
	\label{fig:abilene_nlsr}
\end{subfigure}
\hfill
\begin{subfigure}{0.326\textwidth}
	\includegraphics[width=\linewidth]{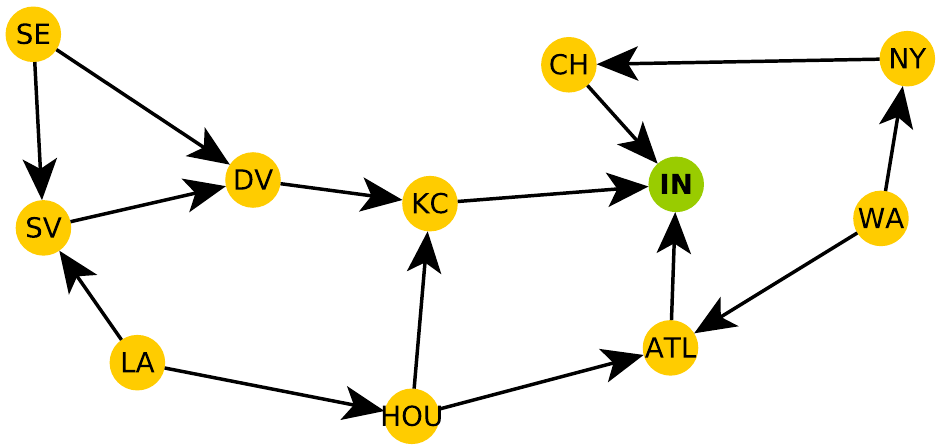}
	\caption{DAG}
	\label{fig:abilene_loopfree}
\end{subfigure}
\hfill
\begin{subfigure}{0.326\textwidth}
	\includegraphics[width=\linewidth]{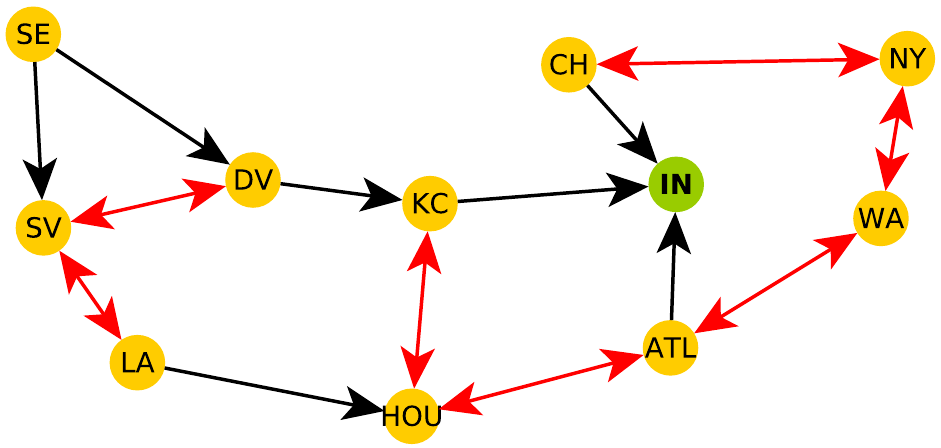}
	\caption{\lfid}
	\label{fig:abilene_nlr}
\end{subfigure}
\caption{Routing entries in the Abilene topology for destination Indianapolis (IN). Comparing strictly Downward Paths, a Directed Acyclic Graph (DAG) using all edges, and Loop-Free Inport-Dependent (\lfid) routing. \vspace*{\figspace}}
\label{fig:abilene_routings}
\end{figure*}

The algorithms above all turn the network into a DAG, with both DWE and MARA achieving the highest nexthop choice theoretically possible in a DAG, since both use one direction for every link for every destination (in contrast, in ECMP and DW, some links are not used for some destinations -- see Figure \ref{fig:abilene_routings}). 
However, DAGs face a theoretical limit in the failure protection they can offer: for every topology and every destination there is at least one link that is not protected against failure (or congestion). This is called the ``last-hop problem'' \cite{iyer2003approach}. Consider the simple triangle and ring topologies in Figure \ref{fig:triangle_ring_dag}.
In a DAG, only node X will be protected against failure of its primary nexthop. 

\section{Moving Beyond DAGs}
\label{sec:alr} \label{sec:nlr} \label{sec:lfid}
\label{sec:design} \label{sec:beyond_dags}

To increase nexthop choice beyond the limits of a DAG, we (per destination) use certain links in both directions.
This leads to many more paths, which can be used to handle both failure and congestion:
Parts of the topology that form a ring can be traversed in both directions, leaving every node with two paths towards the destination (see Figure \ref{fig:triangle_ring_lfip}).
And resilience to single link failure in the Abilene topology (Figure \ref{fig:abilene_routings}) increases from around 30\% to almost 100\% (see Figure \ref{fig:eval_resilience_single} in Section \ref{sec:evaluation}).
However, when using links in both directions, one needs to be careful to avoid loops. We do so with two mechanisms:

First, we exclude 1-hop loops at the data plane. 
Every router will always exclude the incoming port of a packet from the viable nexthop set -- hence the name \emph{Loop-Free Inport-Dependent} routing.
For example, if node Atlanta (ATL) in Figure \ref{fig:abilene_routings} receives a packet from Houston (HOU), it will only consider Indianapolis (IN) and Washington (WA) as nexthops, but never send the packet back to Houston.

Second, one needs to avoid loops longer than one hop. Preferably, while also maximizing link \& node protection and minimizing path stretch. 
Unfortunately, there is no simple rule like the downward criterion to do this. 

Thus, we approach the problem as follows. For each destination: 
First, we add all nexthops, distinguishing between ones going closer to the destination (downward) and ones moving further away (upward) -- see Section \ref{sec:design_add}. 
Second, we iterate through all upward nexthops and remove the ones that would cause a loop (Section \ref{sec:design_loop_removal}). 
Here, the ordering in which nexthops are checked is crucial. We discuss the one that produces the best results in Section \ref{sec:design_order}.
Lastly, this loop removal process can leave certain nodes as a dead end, where incoming packets can only return to the previous node. We prune these dead ends in the final step (Section \ref{sec:design_remove_de}).

\subsection{Adding Nexthops \& Deciding Their Cost\hspace*{-.3em}} 
\label{sec:design_add}

\begin{figure}
\centering
\begin{subfigure}{0.23\linewidth}
\includegraphics[width=\linewidth]{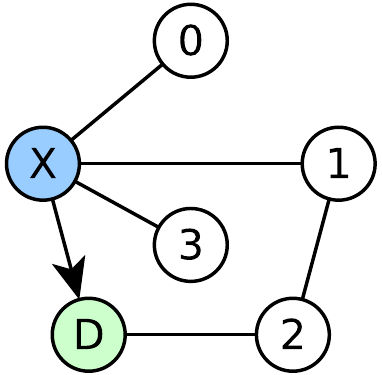}
\caption{}
\end{subfigure}
\hfill
\begin{subfigure}{0.2\linewidth}
\includegraphics[width=\linewidth]{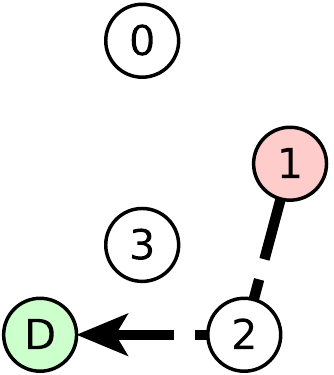}
\caption{}
\end{subfigure}
\hfill
\begin{subfigure}{0.2\linewidth}
\includegraphics[width=\linewidth]{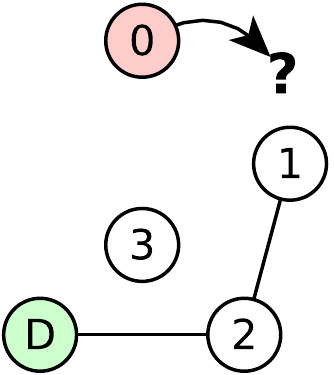}
\caption{}
\label{fig:heuristics_c}
\end{subfigure}
\hfill
\begin{subfigure}{0.2\linewidth}
\includegraphics[width=\linewidth]{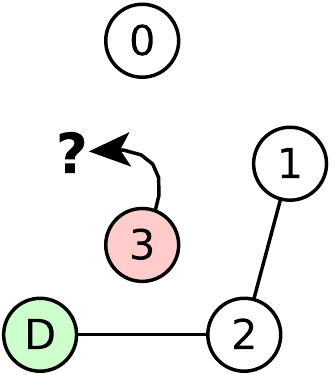}
\caption{}
\label{fig:heuristics_d}
\end{subfigure}

\caption{Avoiding obvious loops. a) base topology; b-d) candidate nexthops 1, 0, and 3. 
\vspace*{\figspace}}
\label{fig:heuristics}
\end{figure}

We assume that each link is given a cost/weight, for example, based on its propagation delay. 
Given this link cost, what cost metric should be assigned to each nexthop in the FIB? 
Intuitively, it should be the cost of the shortest path a packet can take by using this nexthop. 
This cost can be computed efficiently by adding the link weight between current node $x$ and nexthop $n_i$ to the shortest path cost \emph{from} the nexthop to the destination: cost($x$ to $dst$ via $n_i$) = w($x$, $n_i$) + sp($n_i$, $dst$).

However, this approach has one drawback: Sometimes a nexthop can only reach the destination by going back through the current node, causing a loop. Consider Figure \ref{fig:heuristics}, where current node X sends packets to destination D. All three nexthops (0, 1, and 3) have the same shortest path cost of 2 hops, but only nexthop 1 can be used without looping back to X.
This flaw can be fixed by calculating the shortest path cost of neighbors in a graph where the current node X (or all of its links) was removed. Now, nexthops that would loop back through X will receive a cost of infinity, thus will not be added to the FIB (see Figures \ref{fig:heuristics_c} and \ref{fig:heuristics_d}).

\begin{algorithm}
\footnotesize
\caption{Filling the FIB for all nodes}
\label{alg:fill_fib}
\SetInd{0.5em}{1.0em}

\Fn{fillFib (Graph g)} {
AllNodeFib fib\;
\ForAll{$x$ in allNodes}{
map$\langle$DstId, Cost$\rangle$ shortestPathCosts $\leftarrow$ runDijkstra($g$,$x$)\;
map$\langle$nbId, map$\langle$DstId, Cost$\rangle \rangle$ neighborCosts\;
Remove node x from graph $g$\;

\ForAll{$n_i$ in neighbors}{
neighborCost[$n_i$] $\leftarrow$  runDijkstra($g$,$n_i$)\;
}
Add node x back to graph $g$\;
\ForAll{dstId in destinations} {
	spCost $\leftarrow$ shortestPathCosts.at(dstId)\;
	\ForAll{$n_i$ in neighborCosts.at(dstId)} {
		totalNCost $\leftarrow$ $n_i$.cost + linkWeight(nodeId, $n_i$.id)\; 			
		\uIf{neighborCost $<$ spCost} {
			fib[x][dstId].add($n_i$.id, totalNCost, DW)\;
		}
		{
		\ElseIf{neighborCost $<\infty$}{
			fib[x][dstId].add($n_i$.id, totalNCost, UW)\;
		}
		}}
}
}
\Return fib\;
}
\vspace*{\figspace}
\end{algorithm}

The pseudocode for adding neighbors, while avoiding obvious loops, is shown in Algorithm \ref{alg:fill_fib}.
For every node in the network, we compute the shortest path (towards all destinations) once for the node x itself and once for all of its neighbors $n_i$ with node x removed from the graph. 
We then add the nexthops to the FIB, distinguishing between \emph{downward} ($n_i$ is closer to the destination than x), \emph{upward} ($n_i$ is further away), and \emph{disabled} ($n_i$ can only reach the destination through x, and this is omitted from the FIB). 

The complexity of the first half of this algorithm is as follows (m = number of links; n = number of nodes; k = number of neighbors per node): $n$ (for all nodes) * $k$ (for all neighbors) * $m+n\log n$ (for dijkstra's algorithm) = $O(kmn + kn^2\log n)$.  
The complexity of the second half is $n$ (for all nodes) * $n$ (for all destinations) * $k$ (for all neighbors) =  $O(kn^2)$.
Thus, the total complexity remains $O(kmn + kn^2\log n)$.

\subsection{Removing Loops} 
\label{sec:loop_removal}
\label{sec:design_loop_removal}

After determining the cost and type of each nexthop, we check for each one whether it will cause a loop, and remove the ones that do.
We only need to check \emph{upward nexthops}, i.e., ones that lead further away from the destination., since each loop contains at least one upward step. 
Thus, after removing all loop-causing upward nexthops, the network is loop-free.

\begin{algorithm}
\footnotesize
\caption{Removing Loops}  %
\label{alg:remove_loops_de}

\SetInd{0.5em}{1.0em}

\Str{NodePrio}{
nodeId, remainingNh, set$\langle$upwardNh$\rangle$\;
}
\Fn{removeLoops (allNodeFib)}{
\ForAll{dstId in destinations} {
DiGraph dg $\leftarrow$ getDigraphFromFib(allNodeFib.at(dstId))\;
// Queue ordered by max. remainingNh, then cost \\
priorityQueue$\langle$NodePrio$\rangle$ pq\;
pq.push(allNodeFib.getAllUwNexthops(dstId))\;

\While{!pq.empty()}{
node $\leftarrow$ pq.pop()\;
nh $\leftarrow$ node.getHighestCostUwNh()\;
// Remove opposite of upward NH from graph: \\
dg.erase(nh.id, node.id)\; 
// Check if Node is still reachable from uwNh: \\
bool willLoop $\leftarrow$ dg.isConnected(nh.id, node.id)\;
\If{willLoop}{
node.remainingNh$--$\;
allNodeFib[nodeId][dstId].remove(nh)\;
dg.erase(node.id, nh.id)\;
}
dg.add(nh.id(), node.id()); // Add opposite link back \\
\If{node.hasRemainingUwNHs()}
{
pq.push(node)\;
}
}
}
}
\vspace*{\figspace}
\end{algorithm}

As shown in Algorithm \ref{alg:remove_loops_de}, we simulate the network graph for each destination (both downward and upward nexthops are arcs in the graph). 
We iterate through all upward nexthops, ordered by using a priority queue (see next subsection), and perform the loop-check as follows:
For each of the upward nexthops ($x \rightarrow n_i$), we temporarily remove the opposite of the upward link ($n_i \rightarrow x$), and then check whether there is a path from $n_i$ to $x$.
If a path exists, it means that the upward nexthop may cause a loop and thus we remove the nexthop ($x \rightarrow n_i$) from the graph and from the FIB. 
If there is no path, it means the nexthop cannot cause a loop, so we move on to the next one in the list.

\begin{figure}
\centering
\begin{subfigure}{0.23\linewidth}
\includegraphics[width=\linewidth]{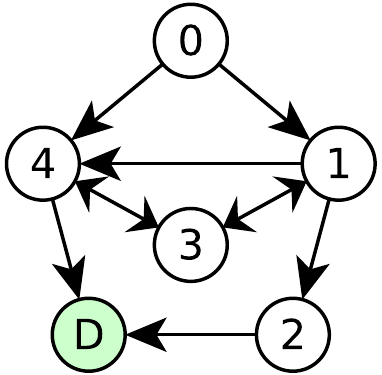}
\caption{}
\end{subfigure}
\hfill
\begin{subfigure}{0.23\linewidth}
\includegraphics[width=\linewidth]{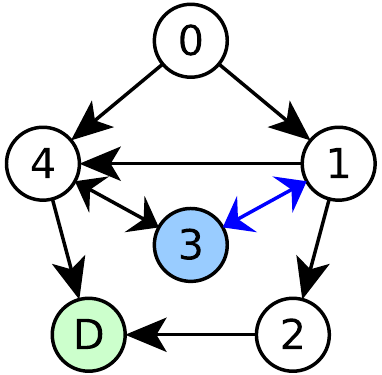}
\caption{}
\end{subfigure}
\hfill
\begin{subfigure}{0.236\linewidth} %
\includegraphics[width=\linewidth]{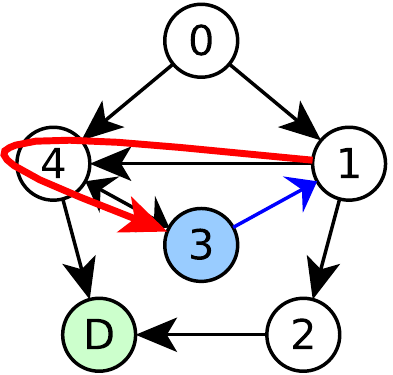}
\caption{}
\end{subfigure}
\hfill
\begin{subfigure}{0.23\linewidth}
\includegraphics[width=\linewidth]{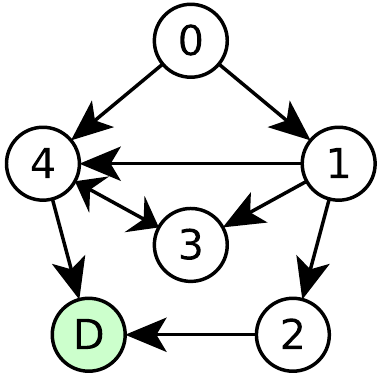}
\caption{}
\label{fig:remove_loops_d}
\end{subfigure}

\caption{Avoiding non-obvious loops. a) base topology; b) candidate upward nexthop (3$\rightarrow$1); c) BFS loop check; d) topology afterwards. \vspace*{\figspace}}
\label{fig:remove_loops}
\end{figure}

We give an example in Figure \ref{fig:remove_loops}. We select upward nexthop  (3$\rightarrow$1) for the check (b). After removing the reverse nexthop,  (1$\rightarrow$3), there still exists a path from 1 to 3 (c), thus we remove upward nexthop (3$\rightarrow$1) from the graph (d).

Since 1) all remaining upward nexthops will not cause loops longer than 1 hop, 2) a sequence of downward nexthops cannot loop, and 3) 1-hop loops are avoided by excluding the incoming port at each step, it follows that every possible path in the network is guaranteed to be loop-free.

\subsection{Ordering of Nexthops}
\label{sec:design_order}

For the loop-removal step, the order of nexthops is crucial: the resulting network is always loop-free, but some orders require more nexthops to be removed and/or lead to longer paths. %
We obtained the best results with the following order: 
\begin{enumerate}
\item Sort all nodes by their number of remaining total nexthops (upward + downard), starting with the node with the most remaining nexthops.
\item If multiple nodes have the same number of remaining nexthops, sort those by the cost of their most costly upward nexthop (starting with the highest).
\end{enumerate}

Both of these steps can be implemented efficiently by using a priority queue filled with an ordered data structure, which we call \emph{NodePrio} (see Alg. \ref{alg:remove_loops_de}). 
The algorithm will pick the first node from the priority queue (which has the most remaining nexthops, and on a tie the most costly upward nexthop), and then pick its most costly upward nexthop for the loop check. 
If the algorithm removes this nexthop, it will reduce the number of remaining total nexthops for the current node, thus changing its position in the queue. 
Lastly, each upward nexthop is only checked once, so nodes without any remaining upward (not just total) nexthops will drop out of the queue.

Nexthops that are checked for loops earlier in this process have a higher chance of being removed, since the graph contains more upward nexthops that can contribute to a loop. 
Thus, ordering nodes by their number of remaining nexthops helps to equalize the number of remaining nexthops after the loop removal is complete.
For example, a node with 4 nexthops (3DW, 1UW) will be checked earlier than another one with only 2 nexthops (1DW, 1UW), leaving the latter one a higher chance of keeping its upward nexthop.
Moreover, checking higher cost upward nexthops before lower cost ones helps to prune longer paths rather than shorter paths, leaving the remaining paths shorter than otherwise.

The complexity of the loop removal is $n$ (for all destinations) * $m$ (upper bound for all upward nexthops) * $m+n$ (for the connectivity check) = $O(mn *(m+n))$ = $O(m^2n)$, since $m > n$.
This is the most time-consuming step in our algorithm.

However, the actual runtime (Section \ref{sec:ev_time}) is much faster than this (worst-case) complexity implies. 
The connectivity check is most efficiently implemented with a bidirectional (BFS) search, which on average runs much faster than m+n. 
For example, in the Sprint topology, only an average of 32 (out of 315) nodes and 46 (out of 972) links need to be visited.

\subsection{Removing Dead Ends}  
\label{sec:remove_deadends}
\label{sec:design_remove_de}

\begin{algorithm}[t]
\footnotesize
\caption{Removing Dead Ends}  %
\label{alg:remove_deadends}

\SetInd{0.5em}{1.0em}
\Fn{removeDeadEnds (allNodeFIB)} {
\ForAll{dstId in destinations} {
	set$\langle$pair$\langle$NodeId, FibNextHop$\rangle$$\rangle$ uwNhSet\;
	uwNhSet $\leftarrow$ allNodeFIB.getUwNhs()\;
	\While{!uwNhSet.empty()}{
		pair$\langle$nodeId, nh$\rangle$ $\leftarrow$ uwNhSet.pop()\;
		// Get reverse FIB entries: \\ 
		reverseEntries $\leftarrow$ {allNodeFIB.at(nh.id).countNh(dstId)}\;
		\BlankLine
		// If there is just one reverse entry: found dead end!\\
		\If{reverseEntries == 1} {
			allNodeFIB.at(nodeId).erase(dstId, nh)\;
			// Push into Set: All NhEntries that lead to nodeId! \\
			uwNhSet.push(allNodeFIB.at(dstId).getNhsTo(nodeId));
		}
	}
}
}
\vspace*{\figspace}
\end{algorithm}

This exhaustive search through all upward nexthops avoids all forwarding loops, but can lead to a small number of \emph{dead ends}, cases where a router receives a packet (via an upward nexthop) but its only forwarding option is to directly return the packet to the previous node.
Fortunately, compared to loops, dead ends are easy to detect and remove (see Algorithm \ref{alg:remove_deadends}): We iterate through all remaining upward nexthop entries and check which of them lead to a node with a FIB size of 1 (meaning that the only nexthop of the neighbor is the downward nexthop leading back to the original node). 
We then remove these upward nexthops, effectively eliminating all dead ends. 

An example of a dead end is shown in Figure \ref{fig:remove_loops_d}. Nexthop (4$\rightarrow$3) leads to a node (3) which can only directly return the packet (FIB size=1), thus this nexthop should be removed. 

The time complexity of this step ($O(mn)$) is smaller than the one of adding nexthops or removing loops.

\subsection{Complexity Analysis}
\label{sec:design_complexity}

Combining the time complexity (m=links; n=nodes; k=neighbors) of the earlier steps, we get $O(kmn + kn^2\log n)$ for adding FIB nexthops + $O(m^2n)$ for the loop removal + $O(mn)$ for the deadend removal = $O(m^2n + kn^2\log n)$. 

\begin{table}
\small
\centering
\caption{Worst Case Time Complexity}
\label{tab:complexity_new}
\begin{tabular}{@{}lll@{}}
	\toprule
	Algorithm & At each node & Network Total \\
	\midrule
	ECMP			& $O(m + n \log n)$ & $O(mn + n^2 \log n)$ \\ 
	DW, DWE		& $O(km + kn \log n)$ & $O(mn + n^2 \log n)$  \\ 
	MARA-MC	\hspace*{-1em}	& $O(mn + n^2)$ & $O(mn + n^2)$ \\ 
	MARA-SPE	\hspace*{-1em}	& $O(mn + n^2 \log n)$ & $O(mn + n^2 \log n)$ \\ 
	LFID \hspace*{-0.8em} & $O(m^2n + kn^2\log n)$ & $O(m^2n + kn^2\log n)$ \\
	\bottomrule
\end{tabular}
\vspace*{\figspace}
\end{table}

A comparison with the related work is shown in Table \ref{tab:complexity_new}.
Similar to MARA \cite{mara2009}, LFID has the same complexity whether it's run for a single node or for the whole network. Thus, it is possible (but not necessary) to compute the routing table once and push it to all routers.
The worst-case time complexity of LFID is higher than MARA by up to a factor of m, the number of links in the network. 
However, as discussed in Sections \ref{sec:design_order} and \ref{sec:ev_time}, the average-case runtime is much closer.

A quick note on \emph{space complexity.} 
In \lfid the router computing its nexthops needs to store the neighbor cost and type (DW, UW) of all nodes, not just its own.
Thus, the space complexity increases from $O(n k)$ for computing downward paths to $O(n^2 k)$. 
If the node id is stored as 2 Bytes (allowing up to 65536 nodes), the cost is stored as 4 Bytes (up to 4.3 billion), and type is stored as 1 Bit, our largest tested topology (n=315, k=6.17) needs around 3.8 Megabytes of memory. 
This should be feasible, given that current routers have memory in the order of tens of Gigabytes.

\section{Evaluation} 
\label{sec:evaluation}

In this section, we compare \lfid against related work.
To emulate the multipath forwarding behavior, we implement a custom C++ simulator, using the Boost graph library for Dijkstra's algorithm and BFS.

We compare 8 topologies
(Table \ref{tab:topologies}) of different size, node degree (Deg), and link metrics: 
1) The Abilene and GEANT topology with link weights set to the geographical distance in miles, rounded to 10 miles (values range from 11 to 224) and
2) the six measured ISP topologies from the Rocketfuel \cite{rocketfuel2002} dataset with their inferred link weights (ranging from 1 to 22.5, in steps of 0.5) \cite{inferring2002}.

\begin{table}
\centering
\small
\centering
\caption{Evaluation Topologies} %
\label{tab:topologies}

\begin{minipage}{.44\linewidth}
\begin{tabular}{@{}lrrr@{}}
	\toprule
	Name &  N & L & Deg \\% & Metric \\
	\midrule
	Abilene 				& 11	& 14 & 2.55 \\% & 1-100 \\
	Geant 					& 27	& 38 & 2.82 \\% & 1-100 \\
	Exodus 		&	79	& 147	& 3.72 \\ %
	Ebone 		&	87  & 161 & 3.70 \\ %
	\bottomrule
\end{tabular}

\end{minipage}
\begin{minipage}{.44\linewidth}
\begin{tabular}{@{}lrrr@{}}
	\toprule
	Name &  N & L & Deg \\% & Metric \\
	\midrule
	Telstra		& 108 & 153 & 2.83 \\ %
	Abovenet	&	141	& 374 & 5.31 \\ %
	Tiscali 	& 161 & 328 & 4.07 \\ %
	Sprint 		&	\hspace*{-1em}315 & 972 & 6.17 \\ %
	\bottomrule
\end{tabular}
\end{minipage}
\vspace*{\figspace}
\end{table}

\subsection{Algorithms \& Scenarios} 
\label{sec:ev_algorithms}

We compare LFID with the multipath routing algorithms discussed in Section \ref{sec:inc_nh_choice},
all of which differ in how they choose the set of nexthops at each router:

\begin{itemize}
\item \textbf{ECMP:} Equal Cost Multi-Path \cite{ecmp_rfc2992} uses the nexthop of the shortest path $n_{sp}$, plus any nexthop $n_{i}$ with the same cost:
$cost(n_i,dst) = cost(n_{sp},dst)$. 

\item \textbf{DW:} Downward paths \cite{rexford2008toward} include the shortest path nexthop plus any nexthop that is closer to the destination than the current node $x$:
$cost(n_i,dst) < cost(x,dst)$.

\item \textbf{DWE:} Downward + Equal Cost Nexthops include all nexthops from DW and, in addition, all nexthops with both an equal cost to the destination and a lower node id.

\item 
\textbf{MARA-MC} \cite{mara2009} creates a Directed Acyclic Graph (DAG) with the specific goal of maximizing the minimum connectivity among all nodes. %
\item \textbf{MARA-SPE} \cite{mara2009} has a similar goal, except that it always includes the shortest path tree in the graph.

\item \textbf{LFID:} Our algorithm, as described in this paper. 

\item \textbf{OPT:} The optimal result based on the network topology constraint. 
This optimum is usually not achievable via loop-free hop-by-hop routing, and serves to show the theoretical limit of the other schemes.
\end{itemize}

Below, we evaluate the presented routing algorithms in 4 different scenarios: \emph{measured runtime} (\ref{sec:ev_time}), resulting \emph{path length \& number of paths} (\ref{sec:ev_paths}), resilience to \emph{single link \& node failures} (\ref{sec:ev_single_failure}), and lastly resilience to \emph{multiple simultaneous link failures and congestion events} (\ref{sec:ev_mult_failures}).

\subsection{Measured Runtime}
\label{sec:ev_time}

\begin{figure}
\centering
\includegraphics[width=\linewidth]{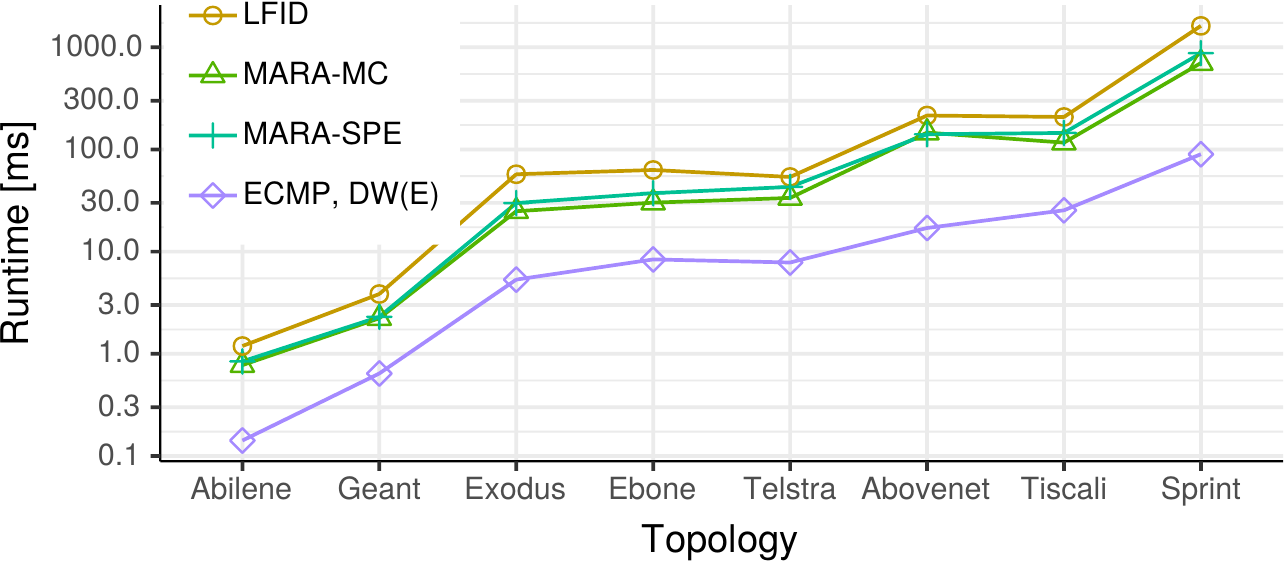}
\caption{Runtime for computing the FIB for all nodes towards all destinations. \vspace*{\figspace}}
\label{fig:eval_time}
\end{figure}

In addition to the complexity analysis in Section \ref{sec:design_complexity}, we measure the actual runtime of the presented algorithms. We ran these measurements on consumer-grade hardware (Intel i7-6600U CPU), single-threaded, calculating the FIB for all nodes towards all destinations. 
As seen in Figure \ref{fig:eval_time}, even though the worst-case complexity implies a much larger difference (roughly 917x higher for the Sprint topology), \lfid is only 25\% to 91\% slower than MARA-SPE.
This is mainly because the loop-removal step (the performance bottleneck) has a much better runtime in the average case than in the worst case (see Section \ref{sec:loop_removal}).  
For larger topologies, there are further options to reduce runtime:

\begin{figure*}
\centering
\includegraphics[width=\linewidth]{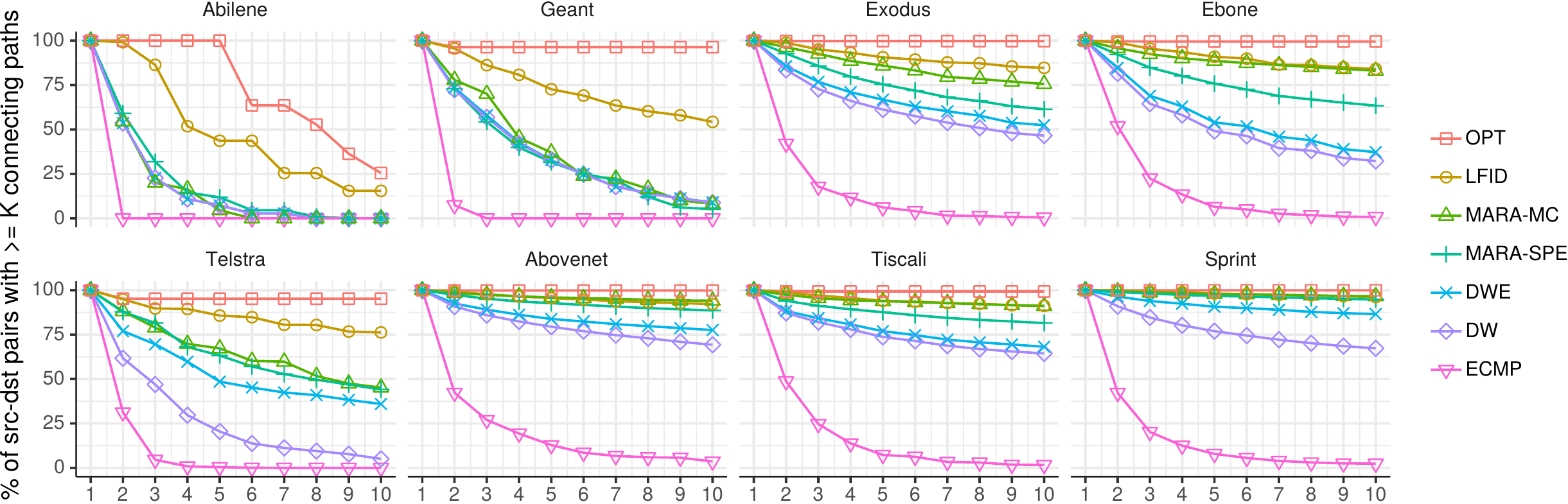}
\includegraphics[width=\linewidth]{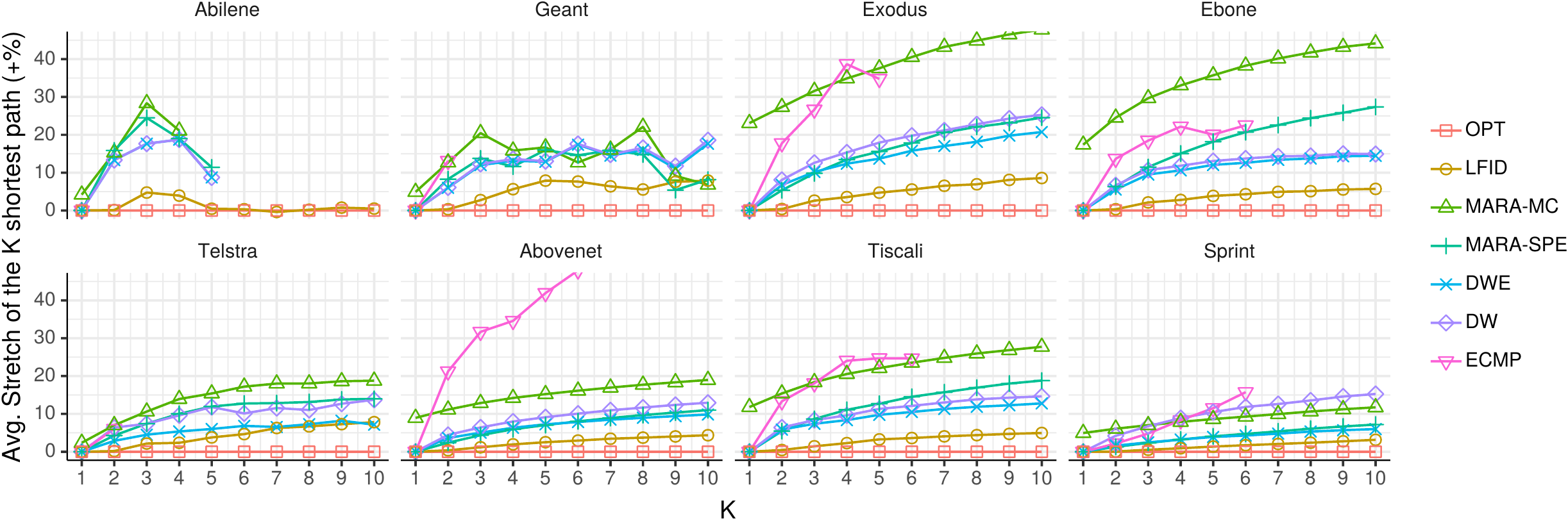}

\caption{Top: Perc. of src-dst pairs that have at least K paths between them; Bottom: Avg. stretch of the K-shortest path.
\vspace*{\figspace}}
\label{fig:ev_k_shortest}
\end{figure*}

First, in contrast to MARA or ECMP, \lfid can easily be parallelized.
The time-critical \emph{removeLoops()} function (Alg. \ref{alg:remove_loops_de}) is run once per destination. 
Since the outcome for each destination does not depend on another, it can be run in parallel, speeding up runtime by a factor of available CPU cores. 

Second, the higher nexthop choice allows instant re-routing through an alternative path, which avoids the need for fast route recomputation during most link failures.
For example, in  88.9\% to 98.2\% of single link failures \lfid can still reach the destination by rerouting at an adjacent node (Section \ref{sec:ev_single_failure}).
Thus, route computation in \lfid is only time-criticial in 1.8\% to 11.1\% of link failures, i.e., 9x--55x less frequent.

\subsection{Number \& Length of Resulting Paths}
\label{sec:ev_paths}

Next, we look at the number of possible paths, and their length (= sum of link costs) provided between each source-destination pair. We do this by running Yen's K-shortest (simple) path algorithm (from \cite{k_shortest_github}) for K=10 paths.
For the optimal result (OPT), we run Yen's algorithm on the \emph{undirected graph} of the base topology. 
For every other scheme, we run Yen's algorithm once for each destination on a \emph{directed graph} that represents the possible nexthops in the FIB. 

Figure \ref{fig:ev_k_shortest} shows the percentage of src-dst pairs that have at least K paths connecting them (top) and their \emph{path stretch} (bottom), i.e., the average ratio of the K-shortest path in the directed vs. undirected topology. 
For K=1 it shows the path stretch of the shortest path, for K=2 the stretch of the second shortest path (if it exists), and so on. 
We removed any path stretch values (bottom plot) where the percentage of paths (top plot) was less than 5\%, since those are likely to be outliers caused by a small sample size. 

Of all the hop-by-hop routing schemes, ECMP, predictably performs the worst. 
In all tested topologies, more than half of source-destination pairs are  connected only through a single path. 
Moreover, ECMP is missing many short paths, thus the average length (stretch) of the K-shortest paths is high.
Downward paths (DW, DWE) are always better than ECMP, and MARA (MC and SPE) is mostly better than downward paths. 
Thus, the ranking according to number of paths is roughly: 
 OPT$>$LFID$>$MARA-MC$>$MARA-SPE$>$DWE$>$DW$>$ECMP.
 
In almost all topologies, \lfid has a higher path choice than all related work.
The exception is the Abovenet topology, where MARA-MC has a higher path choice.
However, MARA-MC buys this higher number of paths with a much higher average path length (Figure  \ref{fig:ev_k_shortest} bottom). Ignoring some outliers of ECMP and DW, MARA-MC's K-shortest paths are the longest of all tested schemes -- up to 50\% longer on average than optimal.
The main reason is that, in contrast to all other schemes, MARA-MC does not always use the shortest path (K=1).
In comparison, MARA-SPE shows a lower path stretch (up to +21\%), but also a significantly lower number of paths.

\lfid shows that one does not have to make the trade-off between high path choice and short potential paths. 
In most topologies, it has the highest number of paths and also the lowest path stretch (less than +9\%). 

However, looking at the K-shortest paths only gives an indirect idea of how many of these paths can be used to circumvent failures and congested links. 
Hence, next we look more directly at the resilience towards single failures/congestions (Section \ref{sec:ev_single_failure}), and at the stretch of the actual paths used by HBH multipath routing (Section \ref{sec:ev_mult_failures}).

\subsection{Resilience to Single Link/Node Failures}
\label{sec:ev_single_failure}

Next, we investigate how this path choice can be used to circumvent a single link failure, link congestion, or node failure. 
Note that for the remaining two experiments, we can treat link failure and congestion interchangeably: 
The question is whether there exists another path that \emph{avoids} a certain link (or multiple links). 
The result is the same for failure and congestion, thus we use the term link/node ``problem'' to denote both cases; we use the term link/node ``protection'' for the ability to circumvent both types of problem.

We consider two different ways of link/node protection: 1) Rerouting adjacent to the problem and 2) rerouting either adjacent or via backtracking to earlier nodes. 

First, we check for an alternative path at the router adjacent to the failure (or congestion). 
This adjacent recovery is easy to implement in practice: After the primary nexthop goes down, a router simply chooses another nexthop from its set.
For all tested algorithms, the resulting path is guaranteed to be loop-free and, in case of a single link failure/congestion, is guaranteed to reach the destination.
It doesn't require signaling between routers, nor backtracking of packets.

\begin{figure}
\centering
\includegraphics[width=\linewidth]{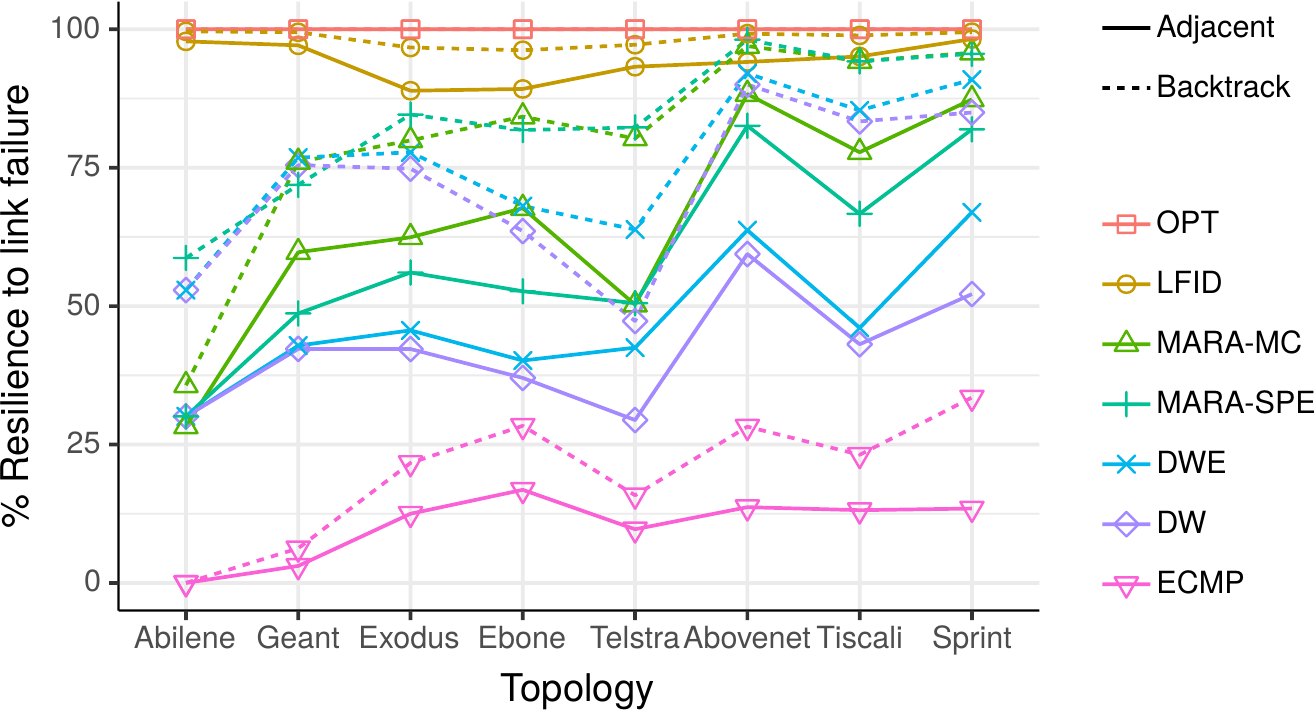}
\includegraphics[width=\linewidth]{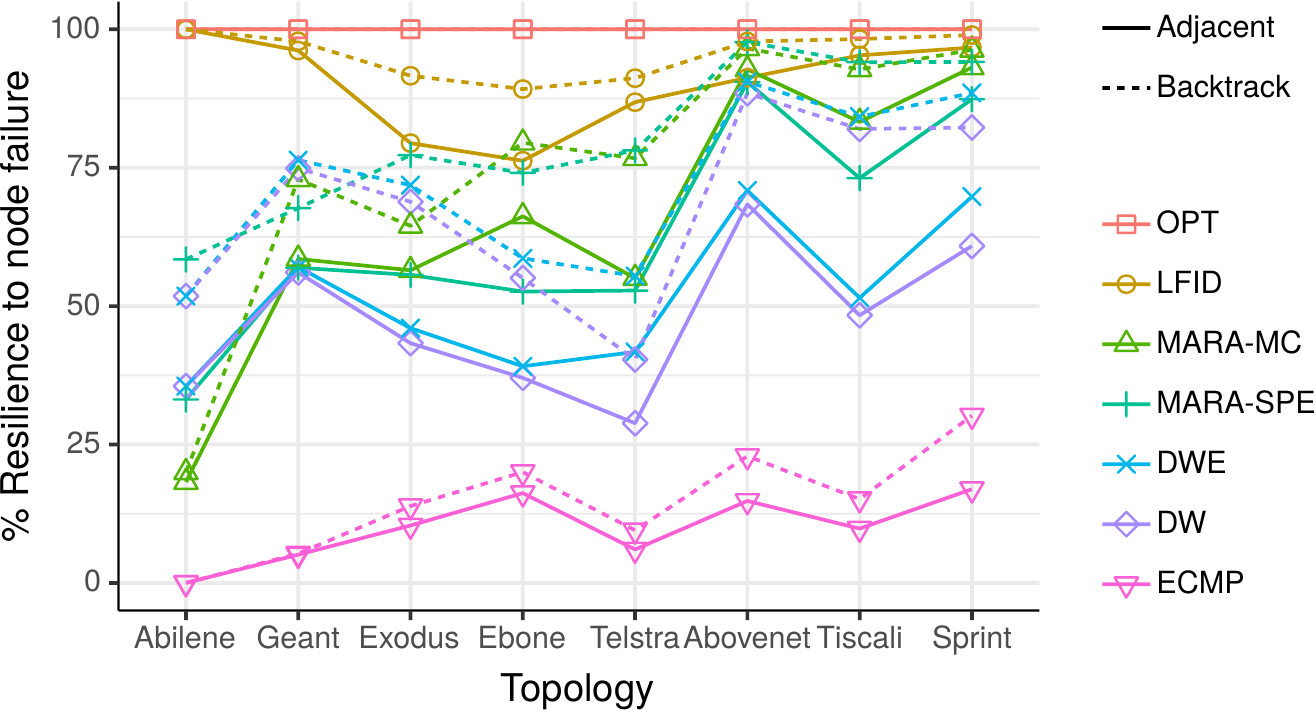}

\caption{Percentage of link/node failures between each src-dst pair that can be recovered at the adjacent node (solid line) or via backtracking (dotted line). \vspace*{\figspace}}

\label{fig:eval_resilience_single}

\end{figure}

Next, we consider the possibility of backtracking on a failure: If the node adjacent to the link/node problem does not have an alternative path/nexthop towards the destination, the packet can be returned to the previous node, which then checks for an alternative path. 
If none is found, the packet will be further backtracked, if necessary, all the way to the source. 
For now, we only investigate the potential link or node protection of such backtracking, and put aside its implementation complexities, such as path stretch, search cost to find a working path, or additional router state.

\begin{figure*}
\centering
\includegraphics[width=\linewidth]{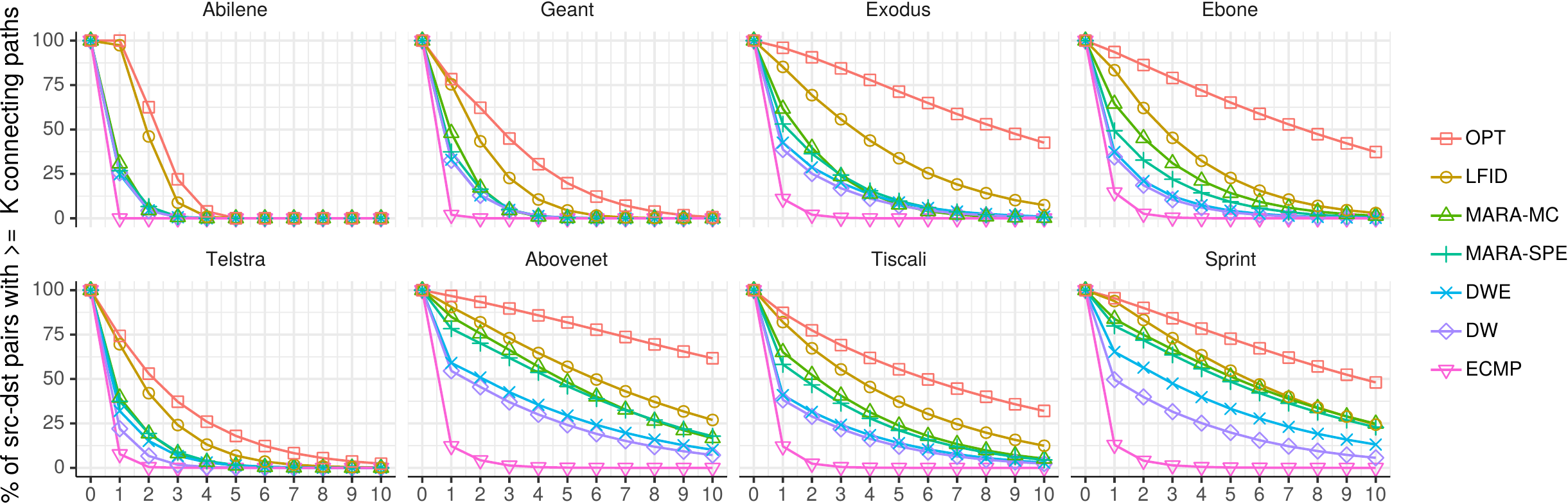}
\includegraphics[width=\linewidth]{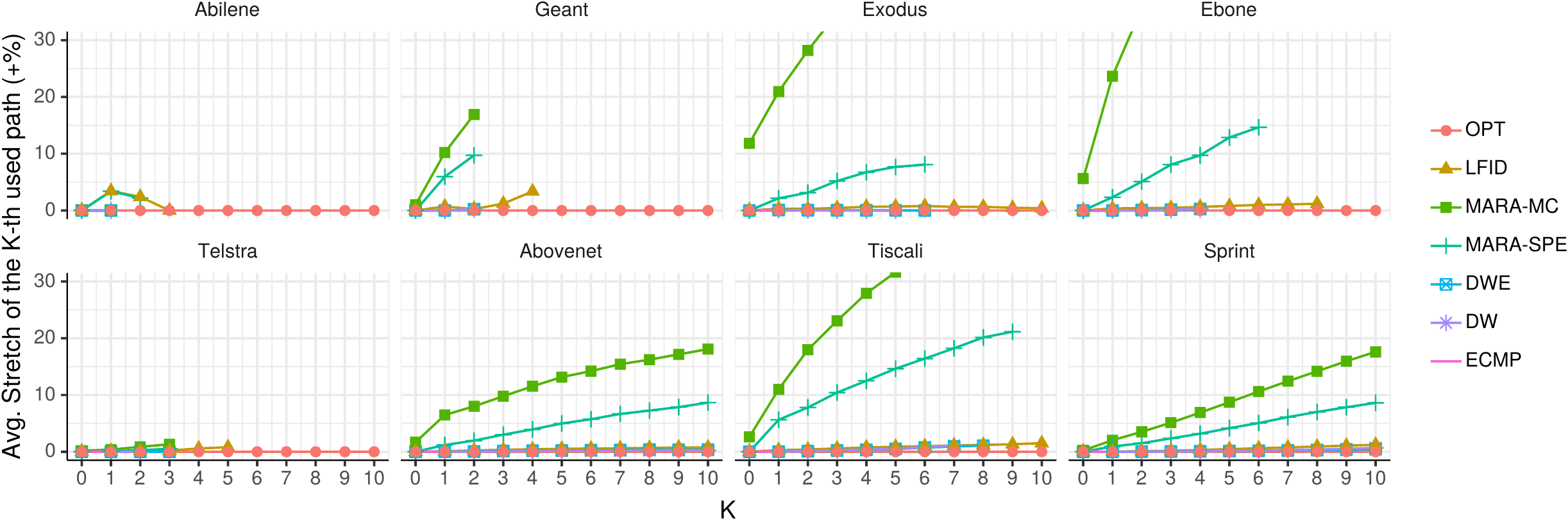}

\caption{Top: Percentage of src-dst pairs that can handle K link problems 
(failures or congestions) 
by rerouting at the adjacent node.
Bottom: Average Stretch of the k-th path used after rerouting.
\vspace*{\figspace}}
\label{fig:ev_mult_cong}
\end{figure*}

More specifically, for the experiment, we look at each link/node on the shortest path between each src-dst pair. 
We remove this link/node and record the percentage of available alternative paths either 1) from the node directly adjacent to the failure, or 2) from the source node (backtracking).
Sometimes, the topology does not contain any possible paths, i.e., when source and destination are disconnected by removing a link or node. 
Thus, we plot the results relative to the maximum protection possible in each topology (OPT), defined as 100\%.

Figure \ref{fig:eval_resilience_single} shows the result for link (top) and node (bottom) protection, where solid lines show adjacent protection and dotted lines protection via backtracking.
For both link \& node failure, and in all tested topologies, LFID outperforms the other tested algorithms\footnote{With a single exception (1 out of 96 data points): For node protection, in the Abovenet topology, and without backtracking, MARA-MC performs slightly better (92.8\%) than LFID (91.2\%).}.
If it seems like other schemes perform better, this is because the plot shows both the adjacent and backtracking case.

LFID can handle 88.9\% to 98.2\% of all recoverable link failures by rerouting at the adjacent node.
Moreover, most of the time\footnote{Except 5 out of 48 data points: Both MARA schemes in the Abovenet topology for both link \& node protection, and MARA-MC in the Ebone topology for node protection.}, LFID provides better link protection \emph{without} backtracking than the other schemes to \emph{with} backtracking. 
Given the complexities of implementing a backtracking scheme (path stretch, search/probing and state overhead), this is an important result.

\subsection{Resilience to Multiple Link Failures \& Congestions}
\label{sec:ev_mult_failures}

Lastly, we evaluate the ability to handle an arbitrary number of link problems (failures or congestions).
Here we focus on \emph{adjacent} protection of failed/congested links. We look at the shortest path between each src-dst pair. 
We remove a \emph{randomly selected link} from the shortest path, then check if there exists another path towards the destination from node adjacent to the removed link (K=1). 
Then, on this second path, we remove another randomly selected link, and check if there exists a third path that avoids both removed links from the new adjacent node to dst (K=2).
And so on, for $K$ = number of avoided links, and $K + 1$ = number of simultaneously used paths if all avoided links are due to congestion.
Note that removed links are specifically chosen to affect a single src-dst pair. Removing 10 random links from a topology will have a much smaller effect on connectivity than K=10, since most of them will not be on the path between a given src-dst pair. 

Moreover, we measure the average stretch of the resulting path, relative to the optimal shortest path that avoids the K removed links.
We plot the average over 100 runs.

As shown in Figure \ref{fig:ev_mult_cong}, again LFID is closer to the optimum resilience than other work (except in the Sprint topology, where it is tied with MARA) providing, compared to the optimal, 69.7\% to 92.5\% resilience against 2 simultaneous failures, and 40.2\% to 86.7\% resilience against 3 simultaneous failures.

Regarding the path stretch, \lfid (and also DW/DWE) performs much better than MARA (MC and SPE): The paths created by adjacent rerouting are on average only 1\% longer than optimal!

\section{Related Work} \label{sec:related_work_back}

\begin{figure}
\footnotesize
\begin{subfigure}{0.38\linewidth}
\includegraphics[width=\linewidth]{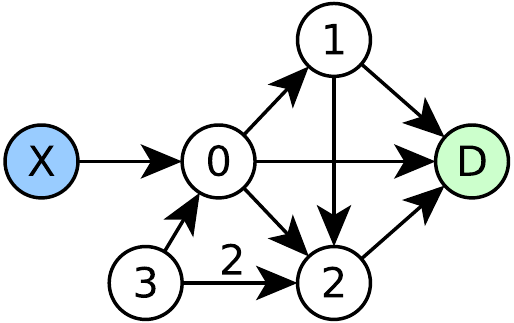}
\caption{DAG}
\label{fig:lfa_dag}
\end{subfigure}
\hspace*{3em}
\begin{subfigure}{0.38\linewidth}
\includegraphics[width=\linewidth]{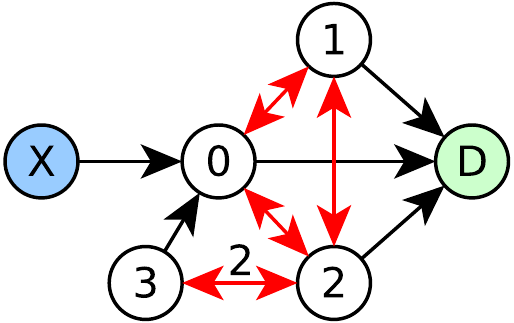}
\caption{LFA}
\label{fig:lfa_lfa}
\end{subfigure}
\caption{Loop-Free Alternate Rules \vspace*{\figspace}}
\label{fig:lfa}
\end{figure}

In addition to the multipath routing schemes discussed in Section \ref{sec:inc_nh_choice}, 
another class of related work is \emph{IP Fast Rerouting} (IPFRR) \cite{ipfrr_rfc}, which provides alternative nexthops for local failure protection on the data plane.
The main difference between IPFRR and the work discussed earlier (ECMP, DW, etc.) is that IPFRR nexthops cannot be used arbitrarily by multiple routers on a path, risking loops when doing so.

A common restriction is that after taking a backup nexthop, packets \emph{must} follow the shortest path. 
Consider the example of Loop-Free Alternates (LFA) \cite{lfa2008,ipfrr2011} in Figure \ref{fig:lfa_lfa}. The alternate nexthops (marked in red) provide protection against any possible link failure, which a DAG cannot \cite{iyer2003approach} achieve  (here: link 2 $\rightarrow$ D is unprotected). 
However, routers cannot freely use LFA nexthops. If they did, packets could form a loop, e.g. 0$\rightarrow$1$\rightarrow$2$\rightarrow$0 or 2$\rightarrow$3$\rightarrow$0$\rightarrow$ 2, and those loops cannot be avoided by excluding the incoming port at each step. 
The restrictions for packets to stay on the shortest path (more generally: use only primary nexthops) severely limits the number of possible paths and thus LFA's ability to deal with multiple simultaneous failures or congestions.
The same reasoning applies to U-turn alternates \cite{uturn-rfc-draft} and IPFRR Tunnels \cite{rfc_not-via}, since those include an even larger set of alternate nexthops.

Another example of IPFRR is permutation routing with joker links \cite{vo2013routing}, which shares following ideas with \lfid: It extends DAGs by using certain links in both directions (there called ``joker links''), and also excludes the incoming link at each hop. 
The main difference is, again, that 1) joker links can only be used when all primary nexthops are down and 2) packets from a joker link can only be sent to a primary nexthop, otherwise risking loops. 
In constrast, in \lfid all routers can use all nexthops simultaneously, even if no links have failed.

Other IPFRR schemes require more sophisticated changes to the IP protocol,
such as per-packet state \cite{dispath2015,elhourani2016ip,robertson2012handling,liu2013ensuring,lakshminarayanan2007achieving} or multi-topology routing \cite{path-splicing2008,kwong2011feasibility, quest2016}. 
In one example of multi-topology routing, Chiesa et al. \cite{quest2016} decompose the routing graph into k arc-disjoint spanning trees. 
When incurring a link failure on its current path, a packet can switch to a different tree, which substantially increases the resilience towards failures. 
However, tree switching incurs a path stretch which, while acceptable for circumventing link failures, is likely unacceptable when splitting up traffic for load balancing, as this would involve many more tree switches. 
In contrast, \lfid ensures that packets rerouted at multiple points stay close to the shortest possible path (see Section \ref{sec:ev_mult_failures}).

Some IPFRR schemes do, however, provide an important benefit: they maintain optimal connectivity for
an arbitrary number of failures \cite{liu2013ensuring,yi2013case}. 
\lfid often comes close to the optimal (see Figure \ref{fig:ev_mult_cong}), but does not reach it.
We leave for future work the question, whether \lfid can be combined with such a data plane mechanism (i.e., dynamic manipulation of FIB state) to ensure optimal connectivity.

\section{Conclusions} \label{sec:conclusion}

We presented \lfid, a simple extension to link-state route calculation, which allows more and shorter loop-free paths than related work.
These paths can be used for either failure protection or congestion reduction (traffic engineering).
We hope that this provides a valuable first step to enable Hop-by-Hop Multipath Routing, where forwarding decisions are made at individual routers inside the network, rather than determined by routers at the edge.

\bibliographystyle{abbrv}
\bibliography{content/bib_new}

\begin{thebibliography}{10}

\bibitem{k_shortest_github}
An implementation of the k-shortest-paths algorithm in cpp.
\newblock https://github.com/yan-qi/k-shortest-paths-cpp-version.

\bibitem{isis_rfc1142}
{OSI IS-IS Intra-domain Routing Protocol}.
\newblock RFC 1142, Feb. 1990.

\bibitem{ecmp_rfc2992}
{Analysis of an Equal-Cost Multi-Path Algorithm}.
\newblock RFC 2992, Nov. 2000.

\bibitem{dispath2015}
S.~Antonakopoulos, Y.~Bejerano, and P.~Koppol.
\newblock Full protection made easy: the dispath ip fast reroute scheme.
\newblock {\em IEEE/ACM ToN}, 2015.

\bibitem{uturn-rfc-draft}
A.~Atlas.
\newblock {U-turn Alternates for IP/LDP Fast-Reroute}.
\newblock Internet-Draft draft-atlas-ip-local-protect-uturn-03, IETF, 2006.

\bibitem{lfa2008}
A.~K. Atlas and A.~Zinin.
\newblock Basic specification for ip fast-reroute: loop-free alternates.
\newblock 2008.

\bibitem{rfc_not-via}
S.~Bryant, S.~Previdi, and M.~Shand.
\newblock {A Framework for IP and MPLS Fast Reroute Using Not-Via Addresses}.
\newblock RFC 6981, Aug. 2013.

\bibitem{quest2016}
M.~Chiesa, I.~Nikolaevskiy, S.~Mitrovi{\'c}, A.~Panda, A.~Gurtov, A.~Maidry,
  M.~Schapira, and S.~Shenker.
\newblock The quest for resilient (static) forwarding tables.
\newblock In {\em IEEE INFOCOM}. IEEE, 2016.

\bibitem{elhourani2016ip}
T.~Elhourani, A.~Gopalan, and S.~Ramasubramanian.
\newblock Ip fast rerouting for multi-link failures.
\newblock {\em IEEE/ACM ToN}, 2016.

\bibitem{segment_routing_rfc8402}
C.~Filsfils, S.~Previdi, L.~Ginsberg, B.~Decraene, S.~Litkowski, and R.~Shakir.
\newblock {Segment Routing Architecture}.
\newblock RFC 8402, July 2018.

\bibitem{fortz2002traffic}
B.~Fortz, J.~Rexford, and M.~Thorup.
\newblock Traffic engineering with traditional ip routing protocols.
\newblock {\em IEEE communications Magazine}, 2002.

\bibitem{fortz2000internet}
B.~Fortz and M.~Thorup.
\newblock Internet traffic engineering by optimizing ospf weights.
\newblock In {\em INFOCOM 2000}, volume~2, pages 519--528. IEEE, 2000.

\bibitem{gojmerac2008towards}
I.~Gojmerac, P.~Reichl, and L.~Jansen.
\newblock Towards low-complexity internet te: the adaptive multi-path
  algorithm.
\newblock {\em Computer Networks}, 2008.

\bibitem{rexford2008toward}
J.~He and J.~Rexford.
\newblock Toward internet-wide multipath routing.
\newblock {\em IEEE Network}, 2008.

\bibitem{iannaccone2004feasibility}
G.~Iannaccone, C.-N. Chuah, S.~Bhattacharyya, and C.~Diot.
\newblock Feasibility of ip restoration in a tier 1 backbone.
\newblock {\em Ieee Network}, 18(2):13--19, 2004.

\bibitem{iyer2003approach}
S.~Iyer, S.~Bhattacharyya, N.~Taft, and C.~Diot.
\newblock An approach to alleviate link overload as observed on an ip backbone.
\newblock In {\em IEEE INFOCOM 2003}.

\bibitem{kvalbein2009multipath}
A.~Kvalbein, C.~Dovrolis, and C.~Muthu.
\newblock Multipath load-adaptive routing: Putting the emphasis on robustness
  and simplicity.
\newblock In {\em ICNP 2009}. IEEE.

\bibitem{kwong2011feasibility}
K.-W. Kwong, L.~Gao, R.~Gu{\'e}rin, and Z.-L. Zhang.
\newblock On the feasibility and efficacy of protection routing in ip networks.
\newblock {\em IEEE/ACM ToN}, 2011.

\bibitem{lakshminarayanan2007achieving}
K.~Lakshminarayanan, M.~Caesar, M.~Rangan, T.~Anderson, S.~Shenker, and
  I.~Stoica.
\newblock Achieving convergence-free routing using failure-carrying packets.
\newblock {\em ACM SIGCOMM CCR}, 37(4):241--252, 2007.

\bibitem{liu2013ensuring}
J.~Liu, A.~Panda, A.~Singla, B.~Godfrey, M.~Schapira, and S.~Shenker.
\newblock Ensuring connectivity via data plane mechanisms.
\newblock In {\em NSDI}, 2013.

\bibitem{inferring2002}
R.~Mahajan, N.~Spring, D.~Wetherall, and T.~Anderson.
\newblock Inferring link weights using end-to-end measurements.
\newblock In {\em IMW}, 2002.

\bibitem{michael2015halo}
N.~Michael and A.~Tang.
\newblock Halo: Hop-by-hop adaptive link-state optimal routing.
\newblock {\em IEEE/ACM Transactions on Networking}, 2015.

\bibitem{path-splicing2008}
M.~Motiwala, M.~Elmore, N.~Feamster, and S.~Vempala.
\newblock Path splicing.
\newblock In {\em ACM SIGCOMM CCR}. ACM, 2008.

\bibitem{ospf_v2_rfc2328}
J.~Moy.
\newblock {OSPF Version 2}.
\newblock RFC 2328, Apr. 1998.

\bibitem{narvaez1999efficient}
P.~Narvaez, K.-Y. Siu, and H.-Y. Tzeng.
\newblock Efficient algorithms for multi-path link-state routing.
\newblock 1999.

\bibitem{mara2009}
Y.~Ohara, S.~Imahori, and R.~Van~Meter.
\newblock Mara: Maximum alternative routing algorithm.
\newblock In {\em INFOCOM 2009}. IEEE.

\bibitem{paganini2009unified}
F.~Paganini and E.~Mallada.
\newblock A unified approach to congestion control and node-based multipath
  routing.
\newblock {\em IEEE/ACM ToN}, 2009.

\bibitem{ipfrr2011}
G.~R{\'e}tv{\'a}ri, J.~Tapolcai, G.~Enyedi, and A.~Cs{\'a}sz{\'a}r.
\newblock Ip fast reroute: Loop free alternates revisited.
\newblock In {\em INFOCOM}. IEEE, 2011.

\bibitem{robertson2012handling}
G.~Robertson and S.~Nelakuditi.
\newblock Handling multiple failures in ip networks through localized on-demand
  link state routing.
\newblock {\em IEEE TNSM}, 2012.

\bibitem{ipfrr_rfc}
M.~Shand and S.~Bryant.
\newblock {IP Fast Reroute Framework}.
\newblock RFC 5714, 2010.

\bibitem{rocketfuel2002}
N.~Spring, R.~Mahajan, and D.~Wetherall.
\newblock Measuring isp topologies with rocketfuel.
\newblock {\em ACM SIGCOMM CCR}, 2002.

\bibitem{ospf-omp-02}
C.~Villamizar.
\newblock {OSPF Optimized Multipath (OSPF-OMP)}.
\newblock Internet-Draft draft-ietf-ospf-omp-02, IETF, Feb. 1999.
\newblock Work in Progress.

\bibitem{mpls_rfc3031}
A.~Viswanathan, E.~C. Rosen, and R.~Callon.
\newblock {Multiprotocol Label Switching Architecture}.
\newblock RFC 3031, Jan. 2001.

\bibitem{vo2013routing}
H.~Q. Vo, O.~Lysne, and A.~Kvalbein.
\newblock Routing with joker links for maximized robustness.
\newblock In {\em IFIP Networking}, pages 1--9. IEEE, 2013.

\bibitem{vutukury1999simple}
S.~Vutukury and J.~Garcia-Luna-Aceves.
\newblock A simple approximation to minimum-delay routing.
\newblock In {\em ACM SIGCOMM CCR}. ACM, 1999.

\bibitem{xu2007deft}
D.~Xu, M.~Chiang, and J.~Rexford.
\newblock Deft: Distributed exponentially-weighted flow splitting.
\newblock In {\em INFOCOM 2007}, pages 71--79. IEEE, 2007.

\bibitem{xu2011link}
D.~Xu, M.~Chiang, and J.~Rexford.
\newblock Link-state routing with hop-by-hop forwarding can achieve optimal te.
\newblock {\em IEEE/ACM ToN}, 2011.

\bibitem{yang2014keep}
B.~Yang, J.~Liu, S.~Shenker, J.~Li, and K.~Zheng.
\newblock Keep forwarding: Towards k-link failure resilient routing.
\newblock In {\em INFOCOM}. IEEE, 2014.

\bibitem{source_selectable_2006}
X.~Yang and D.~Wetherall.
\newblock Source selectable path diversity via routing deflections.
\newblock In {\em ACM SIGCOMM CCR}. ACM, 2006.

\bibitem{yi2013case}
C.~Yi, A.~Afanasyev, I.~Moiseenko, L.~Wang, B.~Zhang, and L.~Zhang.
\newblock A case for stateful forwarding plane.
\newblock {\em Elsevier}, 2013.

\end{thebibliography}

\end{document}